\documentclass[runningheads]{llncs}
\usepackage{epstopdf}
\usepackage{tabularx,threeparttable}
\usepackage{array}
\usepackage[hidelinks]{hyperref}
\usepackage{graphicx}
\usepackage{subfigure}
\usepackage{amssymb, amsmath, bm}
\usepackage{mathtools}
\usepackage{mathrsfs}
\usepackage{booktabs}
\usepackage{multirow}
\usepackage{color}
\usepackage{bbding}

\begin{document}

\title{TransCT: Dual-path Transformer for Low Dose Computed Tomography}
\titlerunning{TransCT}
\authorrunning{Z. Zhang \textit{et al.}}
%
\author{Zhicheng Zhang\inst{1}\inst{(}\Envelope\inst{)} \and Lequan Yu\inst{1,2} \and Xiaokun Liang\inst{1} \and Wei Zhao\inst{1} \and Lei Xing\inst{1}}
\institute{Department of Radiation Oncology, Stanford University, Stanford, USA. \email{zzc623@stanford.edu} \and Department of Statistics and Actuarial Science, The University of Hong Kong}
\maketitle

\begin{abstract}
	
	Low dose computed tomography (LDCT) has attracted more and more attention in routine clinical diagnosis assessment, therapy planning, \textit{etc.}, which can reduce the dose of X-ray radiation to patients. 
	However, the noise caused by low X-ray exposure degrades the CT image quality and then affects clinical diagnosis accuracy. 
	In this paper, we train a transformer-based neural network to enhance the final CT image quality. 
	To be specific, we first decompose the noisy LDCT image into two parts: high-frequency (HF) and low-frequency (LF) compositions.
	Then, we extract content features ($X_{L_c}$) and latent texture features ($X_{L_t}$) from the LF part, as well as HF embeddings ($X_{H_f}$) from the HF part.
	Further, we feed $X_{L_t}$ and $X_{H_f}$ into a modified transformer with three encoders and decoders to obtain well-refined HF texture features.
	After that, we combine these well-refined HF texture features with the pre-extracted $X_{L_c}$ to encourage the restoration of high-quality LDCT images with the assistance of piecewise reconstruction.
	Extensive experiments on Mayo LDCT dataset show that our method produces superior results and outperforms other methods.

\end{abstract}

\section{Introduction}
Computed tomography (CT) system, as noninvasive imaging equipment, has been widely used for medical diagnosis and treatment~\cite{seeram2015computed,mathews2017review}. 
However, concerns about the increase of X-ray radiation risk have become an unavoidable problem for all CT vendors and medical institutions~\cite{brenner2007computed}.
Since x-ray imaging is mainly based on a photon-noise-dominated process~\cite{xu2012low}, lowering the X-ray dose will result in degraded CT images.
Therefore, on the premise of ensuring CT image quality, how to reduce the X-ray radiation dose as far as possible becomes a promising and significant research topic~\cite{brenner2007computed}. 


Compared to sparse or limited-view CT~\cite{zhang2018sparse} and other hardware-based strategies~\cite{zhang2016novel}, 
lowering single X-ray exposure dose~\cite{he2018optimizing,tian2011low} is the most convenient and affordable method.
To obtain high-quality LDCT images, previous works can be mainly classified into two categories: model-based and data-driven methods.
The key to model-based methods is to use a mathematical model for the description of each process of CT imaging:  noise characteristics in the sinogram domain~\cite{manduca2009projection,yu2008sinogram}, image prior information in the image domain, such as sparsity in gradient domain~\cite{laroque2008accurate} and low rank~\cite{cai2014cine}, as well as defects in CT hardware systems~\cite{Zhang2021Modularized}.
This kind of methods are independent of a large training dataset, while the accuracy of the model depiction limits its performance.

With the development of deep learning in medical image reconstruction and analysis~\cite{yu2020deep}, 
many data-driven works have been proposed to reconstruct LDCT images with convolution neural network (CNN)~\cite{wang2020deep}.
Kang \textit{et al.} proposed a CNN-based neural network with the assistance of directional wavelets, suggesting the potential of deep learning technique in LDCT.
Similarly, Chen \textit{et al.} employed residual learning to extract noise in the LDCT images and obtain superior performance~\cite{chen2017low}.
However, these methods need FBP-reconstructed LDCT images as the inputs, which belong to image post-processing.
To get rid of the influence of traditional analytic algorithms (\textit{e.g.} FBP), Zhu \textit{et al.} suggested that `AUTOMAP'  was a direct reconstruction method from the measurement data to the final image~\cite{zhu2018image}.
Then again, the first fully-connected layer as domain transform has a huge memory requirement, which makes AUTOMAP unavailable for large-scale CT reconstruction ~\cite{wang2018image}.
Besides, many works with the combination of iterative reconstruction and deep learning have been proposed as deep unrolled approaches.
This kind of method used CNNs as special regularizations plugged into conventional iterative reconstruction.
They not only inherit the advantages of the convenient calculation of system matrix in conventional algorithms but also get rid of the complicated manual design regularization~\cite{gupta2018cnn,chun2019bcd,he2018optimizing}.
%

Despite the success of CNNs in LDCT reconstruction, CNN-based methods heavily rely on cascaded convolution layers to extract high-level features since the convolution operation has its disadvantage of a limited receptive field that only perceives local areas. 
Moreover, this disadvantage makes it difficult for CNN-based methods to make full of the similarity across large regions~\cite{wang2018non,zhang2019self}, which makes CNN-based methods less efficient in modeling various structural information in CT images~\cite{li2020sacnn}.
To overcome this limitation, Transformers~\cite{vaswani2017attention}, which solely depend on attention mechanisms instead, have emerged as a powerful architectures in many fields, such as natural language processing (NLP)~\cite{devlin2018bert}, image segmentation~\cite{chen2021transunet},image recognition~\cite{dosovitskiy2020image}, \textit{etc.}
In addition to these high-level tasks, Transformer has also been tentatively investigated for some lower-level tasks~\cite{yang2020learning,chen2021pre}, which can model all pairwise interactions between image regions and capture long-range dependencies by computing interactions between any two positions, regardless of their positional distance.

For image denoising, noise is mainly contained in the high-frequency sub-band.
Moreover, the remaining low-frequency sub-band not only contains the main image content, but also contains the weakened image textures, which are noise-free.
These weakened image textures can be used to help noise removal in the high-frequency sub-band.
Inspired by this observation, in this paper, we present the first work, TransCT, to explore the potential of transformers in LDCT imaging.
Firstly, we decompose the noisy LDCT image into high-frequency (HF) and low-frequency (LF) parts.
To remove the image noise on the premise of retaining the image content, we extract the corresponding content features ($X_{L_c}$) and latent texture features ($X_{L_t}$) from the LF part.
Simultaneously, we extract the corresponding embeddings ($X_{H_f}$) from the HF part.
Since transformers can only use sequences as input, we then convert $X_{L_t}$ and $X_{H_f}$ into separated sequences as the input of transformer encoder and decoder, respectively. 
To preserve the fine details of the final LDCT images, we integrate the output of the transformer decoder and some specific features from the LF part and then piecewise reconstruct high-quality and high-resolution LDCT images by stages.
Extensive experiments on Mayo LDCT dataset demonstrate the superiority of our method over other methods.

\begin{figure}[t]
	\centering
	\centerline{\includegraphics[width=120 mm]{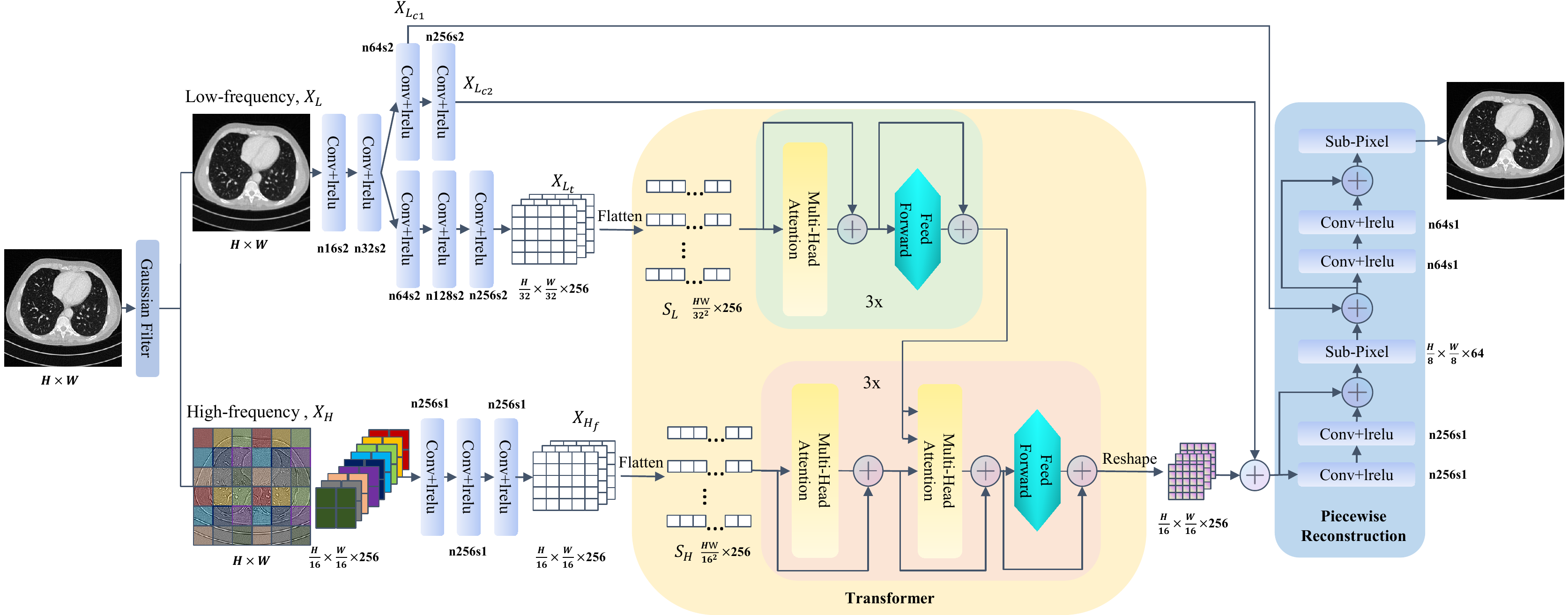}}
	\caption{The overall architecture of the proposed TransCT. 
		`$n64s2$' means the convolution layer has 64 kernels with stride 2.
		Sub-Pixel layer is the upsampling layer~\cite{shi2016real}.
	}
	\label{fig:flowchart}
	\centering
\end{figure}

\section{Method}
\label{sec:method}

Fig~\ref{fig:flowchart} illustrates the overview of our proposed framework.
For image denoising, an intuitive solution is to decompose the noisy image into HF and LF parts, and then the noise is mainly left in the HF part, which also contains plenty of image textures.
However, noise removal only in the HF part breaks the relationship between the HF and LF parts since there are also weakened latent textures in the LF part with reduced noise. 
Therefore, we can remove the noise in the HF part with the assistance of the latent textures from the LF part.
In this work, given the noisy LDCT image $X$ with the size of $H\times W$,  we first use a Gaussian filter with a standard deviation of $1.5$ to decompose the LDCT image into two compositions: HF part $X_H$ and LF part $X_L$.

\begin{equation}
	\begin{aligned}
		X = X_H + X_L	
		\label{X}
	\end{aligned}
\end{equation}

To use the latent textures in $X_L$, we firstly extract the corresponding content fetatures $X_{L_{c}}$ and texture features $X_{L_{t}}$ from $X_L$ using shallow two CNNs. 
Further, we use these texture features and embeddings from $X_H$ to train a transformer and get high-level features of $X_H$, combined with content features from $X_L$ to reconstruct the final high-quality LDCT image.

\subsection{TransCT}

\subsubsection{Sequence}

Similar with what other works have done~\cite{chen2021transunet}, we firstly employ two convolution layers with stride 2 to obtain low-resolution features from $X_L$, and then set two paths to extract content features $X_{L_{c_1}}$ ($\frac{H}{8} \times \frac{W}{8} \times 64$), $X_{L_{c_2}}$ ($\frac{H}{16} \times \frac{W}{16} \times 256$) and latent texture feature $X_{L_t}$ ($\frac{H}{32} \times \frac{W}{32} \times 256$), respectively.
For $X_H$, we employ sub-pixel layer to make $X_H$ to be low-resolution images ($\frac{H}{16} \times \frac{W}{16} \times 256$), and final high-level features $X_{H_f}$ can be obtained with three convolution layers.
The goal is to get a sequence of moderate dimensions eventually.
To take advantage of the characteristic of long-range dependencies of transformers, we perform tokenization by reshaping  $X_{L_t}$ and $X_{H_f}$ into two sequences $S_L$, $S_H$, respectively.

\subsubsection{Transformer}

In this work, we employ a modified transformer with three encoders and three decoders, each encoder includes a multi-head attention module (MHSA) and a feed-forward layer (MLP) and each decoder consists of two multi-head attention modules and a feed-forward layer, as can be seen in Fig~\ref{fig:flowchart}.
For transformer encoder, we use $S_L$ ($\frac{WH}{1024} \times 256$) as the input token, followed by a multi-head attention module to seek the global relationship across large regions, and then we use two fully-connected layers (whose number of the node are $8c$ and $c$, respectively. $c$ is the dimension of the input sequence) to increase the expressive power of the entire network.
\begin{equation}
	\begin{aligned}
		&Z=  MHSA(S_L^{i-1}) + S_L^{i-1} \\
		&S_L^i = MLP(Z) + Z \\
		&s.t.\qquad  i \in \{1,2, 3\}  	
		\label{X}
	\end{aligned}
\end{equation}

After acquiring the latent texture features $S_L^3$ from $X_L$, we feed $S_{H}$ ($\frac{WH}{256} \times 256$) into the first multi-head attention module and treat $S_L^3$ as the key and value of each transformer decoder in the second multi-head attention module.

\begin{equation}
	\begin{aligned}
		&Z =  MHSA(S_{H}^{i-1}) + S_{H}^{i-1} \\
		&Z =  MHSA(Z,S_L^3,S_L^3) +Z \\
		&S_H^i = MLP(Z) + Z \\
		&s.t.\qquad  i \in \{1,2,3\} 	
		\label{X}
	\end{aligned}
\end{equation}

\subsubsection{Piecewise Reconstruction}
Since the transformer only output features $Y$, we combine $Y$ with $X_{L_{c_1}}$, $X_{L_{c_2}}$ to piecewise reconstruct the final high-quality LDCT images.
In our work, the output of the transformer has the size of  $\frac{H}{16} \times \frac{W}{16} \times 256$.
Here, we reconstruct the high-resolution LDCT image piecewise.
In the first step, we add $Y$ and $X_{L_{c_2}}$ and then feed the output into a ResNet with two `Conv2d + Leaky-ReLU(lrelu)' layers, followed by a sub-pixel layer which results in higher-resolution features with size of  $\frac{H}{8} \times \frac{W}{8} \times 64$.
Similarly, we add these higher-resolution features and $X_{L_{c_1}}$.
After another ResNet with two `Conv2d + lrelu' layers and sub-pixel layer, we can get the final output with the size of $H \times W$

\subsection{Loss Function}
The MSE measures the difference between the output and normal dose CT images (NDCT), which reduces the noise in the input LDCT images. Formally, the MSE is defined as follows:

\begin{equation}
	\begin{aligned}
		\min_{\theta} \mathcal L = || I_{ND}-F_{\theta}(I_{LD})||_2^2
		\label{X}
	\end{aligned}
\end{equation}

Where $I_{ND}$ is the NDCT image and $I_{LD}$ is the LDCT image, $F$ is the proposed model and $\theta$ denotes the network parameters.

\subsection{Implementation}
In this work, the proposed framework was implemented in python based on Tensorflow~\cite{abadi2016tensorflow} library. 
We used the Adam~\cite{kingma2014adam} optimizer to optimize all the parameters of the framework.
We totally trained 300 epochs with a mini-batch size of 8.
The learning rate was set as 0.0001 in the first 180 epochs and then reduced to 0.00001 for the next 120 epochs.
The configuration of our computational platform is Intel(R) Core(Tm) i7-7700K CPU @4.20GHZ, 32 GB RAM, and a GeForce GTX TITAN X GPU with 12 GB RAM.
We initialized all the variations with xavier initialization. 
Our code is publicly available at  https://github.com/zzc623/TransCT

\section{Experiments}
\label{sec:experiment}

\subsubsection{Datasets}

In this work, we used a publicly released dataset for \textit{the 2016 NIH-AAPM-Mayo Clinic Low-Dose CT Grand Challenge}\footnote{\url{https://www.aapm.org/GrandChallenge/LowDoseCT/}}~\cite{mccollough2017low}.
In this dataset, normal-dose abdominal CT images of 1$mm$ slice thickness were taken from 10 anonymous patients and the corresponding quarter-dose CT images were simulated by inserting Poisson noise into the projection data. 
To better train the proposed TransCT, we divided the original 10 training patient cases into 7/1/2 cases, related to the training/validation/testing datasets, respectively.
Before network training, we converted CT value of each pixel into its corresponding attenuation value under the assumption that the x-ray source was monochromatic at 60 keV.

\subsubsection{Comparison with other methods}
We compared our method with baseline methods: Non-local Mean (NLM), RED-CNN~\cite{chen2017low}, MAP-NN~\cite{shan2019competitive}, which are the high-performance LDCT methods.
NLM can be found in the scikit-image library\footnote{\url{https://scikit-image.org/}}.
Since there is no public well-trained model for RED-CNN and MAP-NN,
we re-train these methods with the same dataset.

\begin{figure}[t]
	\centering
	\centerline{\includegraphics[width=120 mm]{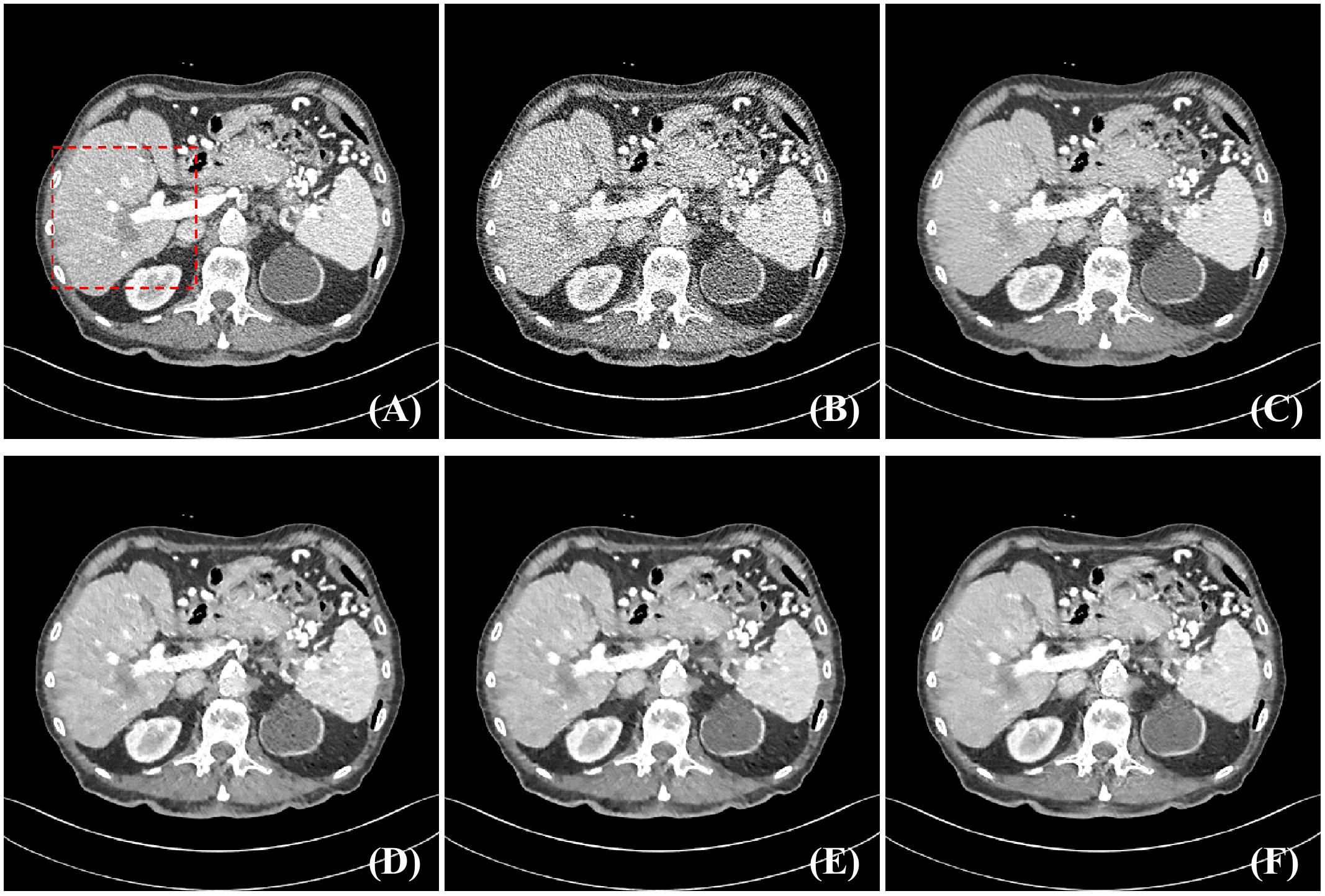}}
	\caption{Visual comparisons from Mayo testing dataset.
		(A) NDCT, (B) LDCT, (C) NLM, (D) RED-CNN, (E) MAP-NN, (F) TransCT. 
		The display window is [-160, 240] $HU$.}
	\label{fig:Figure_2}
	\centering
\end{figure}
\begin{figure}[htb]
	\centering
	\centerline{\includegraphics[width=80 mm]{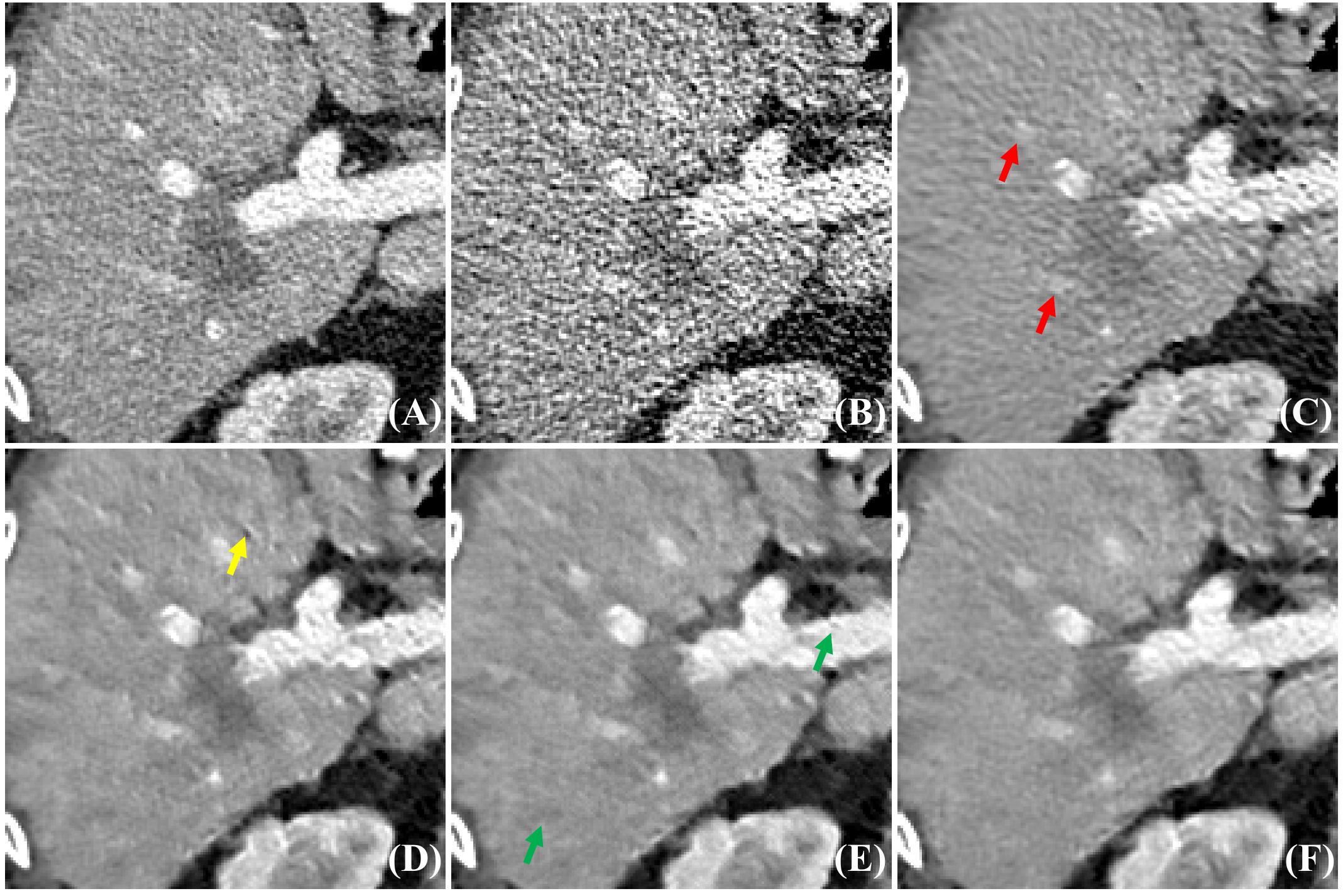}}
	\caption{The zoomed regions marked by the red box in Fig.~\ref{fig:Figure_2} (A). 
		(A) NDCT, (B) LDCT, (C) NLM, (D) RED-CNN, (E) MAP-NN, (F) TransCT.
		The display window is [-160, 240] $HU$.}
	\label{fig:Figure_3}
	\centering
\end{figure}

Fig~\ref{fig:Figure_2} shows the results randomly selected from the testing dataset.
As compared to LDCT (Fig~\ref{fig:Figure_2} (B)), NLM and all the DL-based methods can remove noise to a certain extent, while our proposed TransCT is more close to NDCT.
By investigating the local region in Fig~\ref{fig:Figure_3}, we can see that the blood vessels (red arrows) are not obvious with NLM in (Fig~\ref{fig:Figure_3} (C)). 
RED-CNN and MAP-NN generate, more or less, some additional light tissues (yellow arrow in (Fig~\ref{fig:Figure_3} (D))) and shadows (green arrow in (Fig~\ref{fig:Figure_3} (E))), respectively.

\subsubsection{Quantitative Analysis}
To quantitatively compare all the related methods, we conducted 5-fold cross-validation for all methods on Mayo dataset and employed Root Mean Square Error (RMSE), Structural Similarity (SSIM), and Visual Information Fidelity (VIF)~\cite{sheikh2006image} as image quality metrics.
Among the three metrics, RMSE and SSIM mainly focus on pixel-wise similarity, and VIF uses natural statistics models to evaluate psychovisual features of the human visual system.
From table~\ref{tab:statistc}, we can see that all the related methods improve the image quality on all three metrics.
To be specific, Red-CNN is superior to MAP-NN at the pixel-wise level while inferior to MAP-NN in terms of VIF.
As compared to LDCT, our TransCT can decrease RMSE by 40.5\%, improve SSIM by 12.3\%, and VIF by 93.7\%.
For clinical evaluation, limited by clinical ethics, we evaluated all the methods on clinical CBCT images from a real pig head.
The tube current was: 80$mA$ for NDCT and 20$mA$ for LDCT. 
From table~\ref{tab:statistc}, our method outperforms others with superior robustness.

\begin{table}[b]
	\caption{Quantitative results (MEAN$\pm$SDs) associated with different methods. Red and blue indicate the best and the second-best results, respectively.  }
	\centering
	\begin{tabular}{@{}ccccccc@{}}
		
		\toprule
		Dataset                   & \multicolumn{1}{l}{} & LDCT      & NLM & RED-CNN & MAP-NN & TransCT \\ \midrule
		\multirow{3}{*}{Mayo}     & {RMSE}  & 37.167$\pm$7.245 & 25.115$\pm$4.54 & \color{blue}{22.204$\pm$3.89} & 22.492$\pm$3.897 & \color{red}{22.123$\pm$3.784}     \\
		& {SSIM}        &   0.822$\pm$0.053 & 0.908$\pm$0.031 & \color{blue}{0.922$\pm$0.025} & 0.921$\pm$0.025 & \color{red}{0.923$\pm$0.024}   \\
		& {VIF} & 0.079$\pm$0.032 & 0.133$\pm$0.037 & 0.152$\pm$0.037 & \color{blue}{0.150$\pm$0.038} & \color{red}{0.153$\pm$0.039} \\ \cmidrule(l){1-7}
		
		\multirow{3}{*}{Pig} & {RMSE}  & 50.776$\pm$3.7          & 42.952$\pm$5.971     & \color{blue}{37.551$\pm$5.334}        & 37.744$\pm$4.883      & \color{red}{36.999$\pm$5.25}        \\
		& {SSIM}        & 0.701$\pm$0.02 & 0.799$\pm$0.043    & \color{blue}{0.861$\pm$0.03}        & 0.86$\pm$ 0.027       &   \color{red}{0.87$\pm$0.029}      \\
		& {VIF} & 0.023$\pm$0.002            & 0.040$\pm$0.004 & \color{blue}{0.066$\pm$0.006} & 0.063$\pm$0.006 & \color{red}{0.069$\pm$0.007} \\ \cmidrule(l){1-7} 
	\end{tabular}
	\label{tab:statistc}
\end{table}

\subsection{Ablation study}

\subsubsection{On the Influence of Piecewise Reconstruction}
In this work, after the output of transformer decoder, we used two resnet blocks and two sub-pixel layers to piecewise reconstruct the high-quality high-resolution LDCT image. 
The goal is to restore image detail more finely.
To evaluate the influence of piecewise reconstruction, we modified the proposed TransCT and removed the piecewise reconstruction.
After the output of the third transformer decoder, we used a sub-pixel layer to directly reconstruct the noise-free high-resolution HF texture, and then we added this HF texture and $X_L$ to obtain the final LDCT image.
Specifically, we have removed six convolution layers, including the path of content extraction ($X_{L_{c1}}$ and $X_{L_{c2}}$) and four convolution layers in the final two resnet blocks.
Fig~\ref{fig:Ablation} (a) shows the RMSE value on the validation dataset at each epoch.
We can see that in about the first 20 epochs, the RMSE from modified TransCT decreases faster since its model scale is smaller than our TransCT, while the convergence was inferior to our TransCT with piecewise reconstruction.

\subsubsection{On the Influence of Model Size}
\begin{figure*}[t]
	\centering
	\centerline{\includegraphics[width=80 mm]{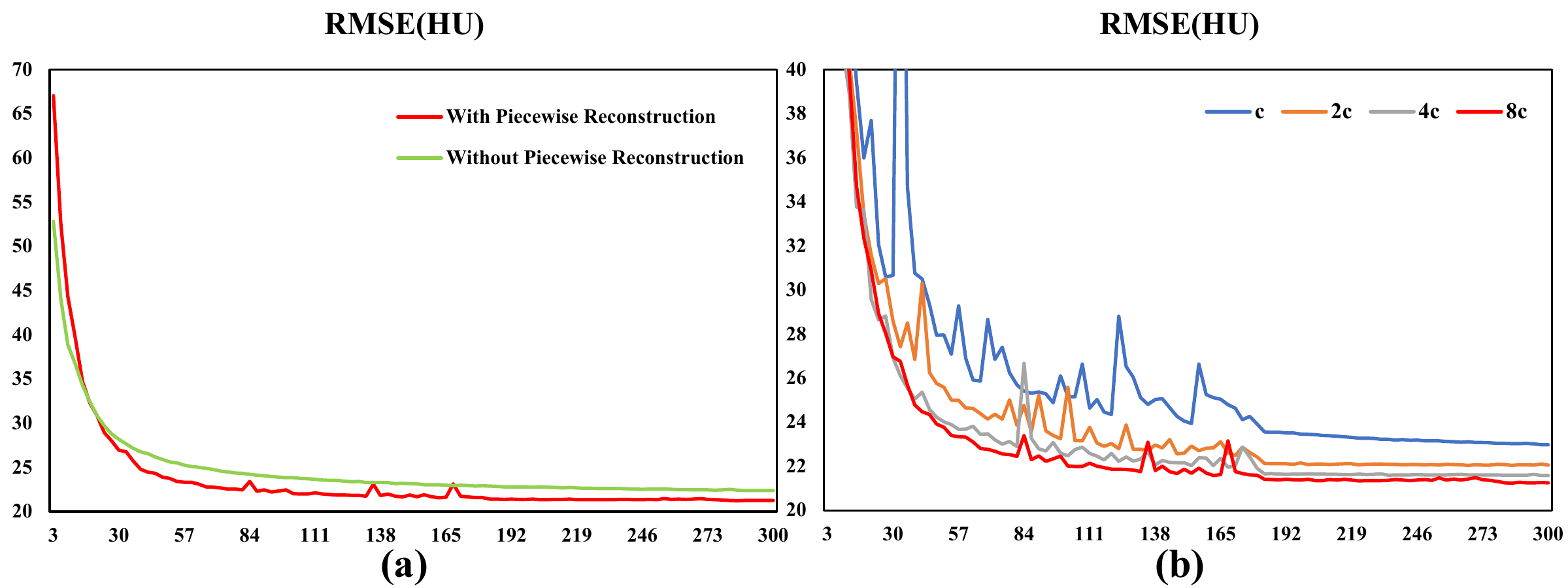}}
	\caption{RMSE results on the validation dataset during the network trainings. }
	\label{fig:Ablation}
	\centering
\end{figure*}

Generally, larger network size will lead to stronger neural network learning ability.
In terms of each transformer encoder and decoder, which includes a two-layer feed-forward network, respectively, when the dimension of the input sequence is fixed, the dimension of the hidden layer in the feed-forward network will determine the network size. 
Here, we adjusted the dimension of the hidden layer \{$c, 2c, 4c$\} to investigate the influence of model size.
From Fig~\ref{fig:Ablation} (b), we can see that the smaller the dimension of the hidden layer is, the larger the fluctuation of the convergence curve is, the larger the final convergent value will be.
Therefore, we conclude that larger model results in a better performance. 
In this work, we set the dimension of the hidden layer in the feed-forward network at $8c$.

\subsubsection{Ablation studies on Transformer Module and Dual-path Module}
To investigate the effectiveness of the transformer module and dual-path module, we conducted  two additional experiments. 
First, we used a revised module ("Conv+3$\times$ResNet blocks") to replace the transformer module. 
We concatenated $X_{H_f}$ and the output from the fourth Conv layer (n128s2, before $X_{L_{t}}$) and then inputted it into the revised module. 
As for the dual-path module, we discarded the HF path and inputted the $X_{L_{c_2}}$ into 3 transformer encoders, whose output will be combined with $X_{L_{c_1}}$ and $X_{L_{c_2}}$ in the piecewise reconstruction stage. 
The results on the validation dataset were shown in table~\ref{tab:ablation}, we can see that our TransCT with transformer module and dual-path module can obtain better performance.

\begin{table}[t]
	\caption{Ablation studies on transformer module and dual-path module conducted on the validation dataset.  }
	\centering
	
	\begin{tabular}{@{}cccc@{}}
		\toprule
		\multicolumn{1}{l}{}                     & RMSE                 & SSIM                 & VIF                  \\ \midrule
		{w/o transformer module} & 22.62 $\pm$2.068 & 0.927$\pm$0.013 & 0.13$\pm$0.023  \\
		{w/o dual-path module}   & 21.711$\pm$1.997 & 0.931$\pm$0.012 & 0.14$\pm$0.025     \\
		{TransCT}                          & 21.199$\pm$2.054 & 0.933$\pm$0.012 & 0.144$\pm$0.025 \\ \bottomrule
	\end{tabular}
	\label{tab:ablation}
\end{table}

\section{Conclusion}
\label{sec:conclusion}
Inspired by the internal similarity of the LDCT image, we present the first transformer-based neural network for LDCT, which can explore large-range dependencies between LDCT pixels.
To ease the impact of noise on high-frequency texture recovery, we employ a transformer encoder to further excavate the low-frequency part of the latent texture features and then use these texture features to restore the high-frequency features from noisy high-frequency parts of LDCT image.
The final high-quality LDCT image can be piecewise reconstructed with the combination of low-frequency content and high-frequency features.
In the future, we will further explore the learning ability of TransCT and introduce self-supervised learning to lower the need for the training dataset.

\subsubsection{Acknowledgements.}
This work was partially supported by NIH (1 R01CA227713) and a Faculty Research Award from Google Inc.

\bibliographystyle{splncs04}

\end{document}